\documentclass[11pt]{article}
\usepackage{graphicx}
\usepackage{listings}
\usepackage{xparse}
\usepackage{seqsplit}
\usepackage{amsmath}
\usepackage{cleveref}
\usepackage{tikz}
\usepackage{pgf}
\usepackage{authblk}

\usepackage{xcolor}   
\usepackage{pgf}      
\usepackage{lmodern}  
\usepackage{graphicx} 
\usetikzlibrary{trees}
\usepackage{listings}

\usepackage{import}
\usepackage{booktabs}
\usepackage{multirow}
\usepackage{pgfplots}
\usepackage{url}
\usepackage{fancyhdr}

\lstdefinestyle{compactcode}{
  basicstyle=\ttfamily\footnotesize, 
  numbers=left,
  numberstyle=\tiny,
  numbersep=6pt,          
  frame=single,
  framesep=2pt,           
  xleftmargin=0pt,        
  aboveskip=4pt,          
  belowskip=4pt,          
  showstringspaces=false,
  columns=fullflexible,   
  breaklines=true,
  breakatwhitespace=true,
  keepspaces=true,
  tabsize=2               
}
\NewDocumentCommand{\icode}{v}{\texttt{\seqsplit{#1}}}

\title{Optimizing Optimism: Up to $3.5\times$ Faster zkVM Validity Proofs via Sparse Derivation}
\author{Mohsen Ahmadvand \quad Pedro Souto}
\affil{Zircuit\\
\texttt{mohsen@zircuit.com} \quad \texttt{pedro@zircuit.com}}
\date{}

\begin{document}
\maketitle

\begin{abstract}
The Optimism derivation pipeline is engineered for correctness and liveness, not for succinct validity proofs. A straightforward port to a zkVM imposes significant overheads, making validity proofs significantly more costly than necessary. We systematically identify inefficiencies in the current design, analyze their impact on proving costs, and provide a soundness-preserving redesign tailored to zk proving. Our redesign achieves up to 6.5× faster derivation inside zkVMs (3.5× overall speedup) while maintaining identical safety guarantees.

\end{abstract}

\section{Introduction}
Zero-knowledge virtual machines (zkVMs)~\cite{sp1-docs} come with memory and cycle limitations. As the input space of the program grows the memory footprint increases. Higher memory needs requires more specialized hardware for the proving infrastructure inuring substantially higher fees. That is, tweaking the program to reduce memory usage and making the program more optimized to consume less cycles achieving the same goal can remove the need to run super specialized hardware.

On the same hand, if the input space can grow beyond the hardware at reach limits or the practical limit of the underlying VMs, a risk of denial of service is imminent (SP1's VM memory is capped at \texttt{0x78000000} bytes by the default allocator). Simply put, if the program that runs inside VM can have an arbitrary input size, there is a point at which no proof can be computed further.
For a safety critical system such a rollup, it is important to identify such unbounded inputs and evaluate the risk of DoS. 

Moreover, rollups and real-time proving go through a constant battle of remaining financially viable. If the same work can be achieved at a lower cost the impact can be significant both on costs and remaining in sync with the latest incoming blocks.

In this paper, we analyze validity proof generation for Optimism~\cite{optimism-specs}-based rollups. We look at each part of the logic constituting the proof time and identify derivation as a part that has the risk of unbounded inputs, and severely optimized logic for zkVM execution.

\section{Optimism Overview}\label{sec:optimism-background}

Optimism is comprised of three primary components: \texttt{op-node}, \texttt{op-batcher}, and \texttt{op-geth}. We refer readers to the official Optimism specifications~\cite{optimism-specs} for a complete review of the system. Here we briefly summarize the system flow at a high level. 

The state of the rollup is summarized in the \emph{output root}, a Merkle root
constructed from the L2 state root, the L2 message passer storage root, the
L1 origin block hash, and the L2 block number. This output root is posted to
L1 and serves as the commitment against which withdrawals are proven.

L1--to--L2 interactions begin through the \textit{Optimism Portal} contract on L1.
Users submit deposits or initiate contract calls via this portal, which emits events that are consumed by \texttt{op-node}. For each inclusion (deposit) transaction,
\texttt{op-node} injects the L1-derived transactions into the next L2 block in a
\emph{fixed, protocol-defined order} ahead of sequenced user transactions.
These L1-derived inclusions are not part of the data-availability batches and
must therefore be applied exactly as they appear on L1.

\texttt{op-geth} runs the execution layer in a permissioned (PoA-style) sequencer setup.
It \emph{orders L2 user transactions}. The forced L1 inclusions provided by
\texttt{op-node} are not reordered by \texttt{op-geth}; blocks are built from
the exact transaction list supplied via payload attributes.

Beyond injecting inclusions, \texttt{op-node} continuously runs the \emph{derivation
pipeline}. The purpose of derivation is to reconstruct the canonical L2 chain
directly from L1 data, allowing any node to sync trustlessly without relying
on the sequencer. This pipeline is triggered whenever new L1 blocks are
produced: \texttt{op-node} scans each block for batcher transactions, validates
their senders against the on-chain \texttt{SystemConfig}, extracts calldata or
blob payloads, and assembles them into frames and channels. In parallel, it
processes Optimism Portal events to inject the corresponding L1→L2
transactions. The resulting ordered stream is then executed by \texttt{op-geth}
to advance the L2 state.

The rollup exposes two canonical heads. The \emph{safe head} reflects L2 blocks
derived from the current L1 head, and can still be reverted if L1 reorgs occur.
The \emph{finalized head} reflects L2 blocks derived only from finalized L1
blocks, which cannot be reverted under normal consensus assumptions. Together,
these notions of safety and finality give nodes both a responsive view of the
rollup’s state and an economically irreversible commitment.

As the L2 state advances, \texttt{op-batcher} collects and compresses sequences
of L2 transactions into batches. Batches are organized into \textit{channels}
with a fixed \textit{channel duration timeout} that bounds how long a channel
may remain open before it must be closed and published to L1, ensuring timely
data availability.

Batches are submitted via batcher transactions on L1. The \textit{BatchInbox}
is a fixed address (not a contract):
\[
\texttt{0xff00000000000000000000000000000000042069}.
\]
For calldata batches, \texttt{op-batcher} sends a normal L1 transaction with
\texttt{to} set to the BatchInbox and the batch bytes in \texttt{calldata}. For
blob batches, the batch bytes live in EIP-4844 blobs referenced by the
transaction (blobs themselves have no \texttt{to} field). In both cases,
\texttt{op-node} monitors L1, authenticates batcher transactions, and ingests
their data for derivation.

The batcher EOA is configured on L1 in \texttt{SystemConfig}. Derivation
authenticates each batcher transaction by checking that \texttt{to} equals the
BatchInbox address and \texttt{from} equals the \emph{current} batch-sender
address from \texttt{SystemConfig} for that L1 block. Accordingly, at the start
of processing every L1 block, derivation refreshes its view of
\texttt{SystemConfig} and applies any batcher EOA updates before accepting
batch data.

\texttt{op-node} continuously monitors L1 for (i) batcher transactions to the
BatchInbox and (ii) Optimism Portal events (deposits). Upon detecting new
batches, it advances the safe head of the L2 chain. In the presence of L1
reorgs, \texttt{op-batcher} may resubmit batches under new \textit{epoch}
numbers; higher epoch numbers for the same time slot take precedence.

\subsection{Optimism Derivation}
Derivation enables any node to rebuild the canonical L2 chain \emph{from L1 alone}, without trusting the live sequencer feed. This trust-minimized path is the foundation for safety (forced inclusions from L1 are respected) and for permissionless synchronization.

Derivation runs continuously as new L1 blocks become available. On each step, the derivation state advances the L1 origin by exactly one block, updates on-chain configuration (e.g., batcher EOA), and exposes batch data (calldata or blobs) to downstream stages.

We outline the exact control flow from the Optimism codebase.

\begin{enumerate}
  \item \textbf{Advance L1 origin and refresh SystemConfig.}
  The traversal stage increments the L1 origin and validates reorg safety in
  \icode{l1_traversal.go/AdvanceL1Block}:
  \begin{itemize}
    \item Fetch next L1 header: \icode|L1BlockRefByNumber(ctx, origin.Number+1)|.
    \item Detect reorgs by checking \icode|ParentHash|.
    \item Fetch receipts: \icode|FetchReceipts(ctx, nextL1Origin.Hash)|.
    \item Update the in-memory \icode|SystemConfig| from receipts via
          \icode{UpdateSystemConfigWithL1Receipts(&sysCfg, receipts, cfg, time)}.
          This captures configuration changes such as the \emph{batcher EOA}.
  \end{itemize}
  This stage also implies: \emph{at the beginning of every L1 block}, derivation
  must respect any configuration changes (e.g., a new batcher EOA) before
  authenticating batcher transactions.
  \item \textbf{Expose batch data to downstream (L1 Retrieval).}
  The retrieval stage pulls per-block data in
  \icode{l1_retrieval.go/NextData}:
  \begin{itemize}
    \item If this is the first request for the current origin, open a data source
          with the \emph{current} batcher address from \icode|SystemConfig|:\\
          \icode|dataSrc.OpenData(ctx, nextL1Block, SystemConfig().BatcherAddr)|.
    \item Iteratively return \icode|Next(ctx)| chunks until EOF for this origin,
          then advance to the next L1 block.
  \end{itemize}
  Operationally: the data source scans the L1 block's transactions and yields
  only valid batcher payloads (calldata or blob pointers).
  \item \textbf{Select calldata vs.\ blobs and authenticate batcher txs.}
  In the blob-aware source, \icode{blob_data_source.go/dataAndHashesFromTxs}
  filters and classifies transactions:
  \begin{itemize}
    \item Validate batcher transactions with \icode{isValidBatchTx(tx, l1Signer, batchInboxAddress, batcherAddr, logger)}.
          This enforces the \textit{BatchInbox} magic address for calldata cases,
          and authenticates the \texttt{from} against the \emph{current}
          batcher EOA.
    \item For non-blob batcher txs: extract \icode|tx.Data()| as batch bytes (calldata path).
    \item For blob txs: collect \icode|tx.BlobHashes()| and create placeholders to be
          filled after blob sidecar download (calldata, if present, is ignored).
  \end{itemize}
  The result is an ordered sequence of \emph{blob-or-calldata} elements for downstream
  channel/frame assembly.
  \item \textbf{Channel / frame assembly and payload attribute generation.}
  Buffer frames from batcher data, assemble complete channels, decode/decompress batches, then produce payload attributes.
\end{enumerate}

\subsection{Complexity Analysis and Cost Drivers}

We analyze the dominant costs per L1 block $B$, with $T_B$ L1 transactions, $R_B$ receipts,
$K_B$ batcher transactions (subset of $T_B$), and $H_B$ total blob hashes referenced by
the batcher transactions in $B$.

\paragraph{AdvanceL1Block (Traversal).}
\begin{itemize}
  \item Header lookup: $O(1)$ (amortized).
  \item Reorg check: $O(1)$.
  \item Receipt fetch + scan: $O(R_B)$ to parse and apply \icode|UpdateSystemConfigWithL1Receipts|.
\end{itemize}
\textbf{Cost drivers:} receipt retrieval/parsing; any config changes force the batcher EOA to be refreshed \emph{per block}
before data authentication.

\paragraph{NextData (Retrieval).}
\begin{itemize}
  \item Opening a data source once per L1 block: $O(1)$ setup, dependent on backend.
  \item Iterative \icode|Next| calls until EOF: proportional to number/size of batch payloads in $B$.
\end{itemize}
\textbf{Cost drivers:} I/O and decoding overhead for large batches; repeated open/close across dense L1 traffic.

\paragraph{dataAndHashesFromTxs (Classification).}
\begin{itemize}
  \item Transaction loop: $O(T_B)$.
  \item For each tx, authentication via \icode|isValidBatchTx| (signature/recipient checks): typically $O(1)$ per tx.
  \item Blob handling: collecting \icode|BlobHashes()| over batcher txs: $O(H_B)$; downloading sidecars later adds I/O proportional to total blob bytes.
\end{itemize}
\textbf{Cost drivers:} scanning \emph{all} L1 txs to find the few batcher txs ($K_B \ll T_B$ on busy chains),
plus blob sidecar retrieval size/latency. Note the code increments a running \icode|blobIndex| even for
non-batcher txs (to preserve correct indexing), which forces inspection across the full block.

\paragraph{Forced inclusions and ordering.}
\begin{itemize}
  \item L1-derived transactions (Portal events) are \emph{not} in DA; they must be read from L1 receipts/logs
        and injected in L1 order: additional $O(R_B)$ scan.
\end{itemize}

\paragraph{End-to-end per L1 block (high level).}
\[
  O\!\left(R_B\right) \;+\; O\!\left(T_B + H_B\right) \;+\; \text{I/O}\_{\text{blobs + calldata}}.
\]
In practice, on mainnet/Sepolia with heavy unrelated L1 traffic, $T_B$ dominates; in blob-heavy regimes $H_B$ and blob
I/O dominate. For zkVM execution of the derivation, these scale factors translate into prover time proportional to:
(i) number of L1 receipts parsed, (ii) number of L1 txs scanned to filter batcher txs, and (iii) total bytes of batch data
downloaded/decoded (calldata or blobs).

\subsection{Observed Inefficiencies (Motivation for Optimizations)}

\begin{itemize}
  \item \textbf{Full-block scanning to isolate batcher txs.} \icode|dataAndHashesFromTxs| inspects \emph{every} L1 tx and
        maintains a global blob index across the block. On busy L1, this produces $O(T_B)$ work per block even when $K_B$ is small.
  \item \textbf{Receipt-driven config updates every block.} \icode|AdvanceL1Block| always fetches and scans receipts to refresh
        \icode|SystemConfig|. Even when the batcher EOA rarely changes, the fixed $O(R_B)$ cost persists.
  \item \textbf{Forced-inclusion ordering path is separate from DA.} Because L1-derived txs are not in DA, derivation must
        \emph{also} parse Portal events from receipts in strict order, adding an additional $O(R_B)$ pass (though often shared with config updates).
  \item \textbf{Channel timeout spillover.} Longer channel duration reduces L1 posting frequency but increases worst-case
        L1 traversal and blob/calldata volume when a channel finally closes, creating bursty loads for the zkVM.
\end{itemize}

\paragraph{Implication.}
Any optimization that (i) reduces unnecessary full-block scans, (ii) amortizes or indexes receipt lookups for
\icode|SystemConfig| and Portal events, or (iii) pipelines blob sidecar retrieval/verification against zkVM execution
will directly lower the asymptotic and constant factors in derivation cost.

\section{Kona: Optimism Fault Proof Program}
Kona~\cite{kona} is a Rust implementation of the OP Stack fault proof program that enables trustless verification of L2 state transitions. Our implementation builds on Kona. Kona's fault-proof program mirrors Optimism's derivation: it boots from on-chain
state, constructs L1/L2/Blob providers, initializes an \texttt{OraclePipeline},
and drives execution via \texttt{Driver::}\allowbreak\texttt{advance\_to\_target}.

\textbf{Problem.} Kona’s faithful replication of Optimism’s core logic also
inherits two dominant costs: (i) scanning every L1 block/tx in the zkVM to
discover DA-bearing batcher transactions, and (ii) fetching and parsing receipts
on every L1 block to detect \texttt{SystemConfig} (batcher EOA) changes.

At a high level, \icode{run(...)} loads \icode{BootInfo}, computes the safe head,
builds providers (\icode{OracleL1ChainProvider}, \icode{OracleL2ChainProvider},
\icode{OracleBlobProvider}), initializes \icode{EthereumDataSource} and the
\icode{OraclePipeline}, and executes with \icode{KonaExecutor}. L1 traversal
advances one L1 block at a time, applies reorg checks, and (in the baseline)
updates \icode{SystemConfig} from receipts; retrieval then exposes DA (calldata
or blobs) to downstream stages which assemble frames and channels and inject
L1-forced inclusions in strict L1 order ahead of sequenced user transactions.

\subsection{ZK Validity Proof Programs}
Building a zk-Optimism begins with the Kona program: given a range of L2 blocks,
we check that the state transitions from a known-good root to the client’s
claimed root.

The \icode{L2OutputOracle} contract tracks the last agreed output root together
with the last L2 block number and corresponding L1 origin (which anchors DA and
deposits). In our setting, \icode{proposeOutputRoot} is extended to accept a
succinct zero-knowledge proof for trustless on-chain verification. We use Groth16~\cite{groth16}, a pairing-based SNARK scheme that produces constant-size proofs with efficient verification, as the proving system for our Optimism-based implementation. While other zk-SNARK schemes could be used without loss of generality, Groth16's balance of proof size and verification cost makes it well-suited for L1 settlement.

The fault-proof system comprises two subprograms: \emph{range} and \emph{aggregate}.
The \emph{range} program takes $(\textit{start}, \textit{end}, \textit{safe\_root},
\textit{claim\_root})$ and retrieves the necessary L1/L2 hints to validate the
claimed transition. The \emph{aggregate} program verifies a sequence of range
proofs for contiguity and computes the terminal root; it outputs a single
\textsc{Groth16} proof submitted to \icode{L2OutputOracle} to authorize
withdrawals against the verified output root.

\section{Optimization: Nonce- and Bloom-Hash–Based Derivation}

Executing Optimism's derivation \emph{verbatim} inside a zkVM forces the prover to (i) traverse \emph{every} L1 block and scan \emph{all} transactions to discover DA-bearing batcher submissions, and (ii) fetch/parse \emph{all} receipts to detect \icode{SystemConfig} updates (notably batcher EOA changes). Our goal is to optimize these searches inside the zkVM while preserving the protocol's trustless, sound behavior. As we show in the following, under certain circumstances, we toggle the optimized path off resorting to the ``baseline'' derivation.

\subsection*{Optimization Approach (prefeed+nonce)}
We optimize the in-VM searches with two complementary mechanisms:
\begin{enumerate}
  \item \textbf{DA-aware prefeeding.} The host precomputes a sparse set $\mathcal{B}$ of L1 block numbers that actually contain DA-bearing batcher transactions (calldata or blobs) and provides \emph{only} those blocks to the zkVM. When there are no \icode|SystemConfig| updates, non-DA blocks are omitted entirely.
  \item \textbf{Bloom-gated config updates.} Rather than parsing receipts every block to detect \icode{SystemConfig} updates, the zkVM first checks the L1 header’s logs-bloom for the batcher-update topic; it fetches and parses receipts \emph{only on bloom hits}. Since blooms have no false negatives for topics, correctness is preserved; false positives merely cause harmless (more on this in~\Cref{sec:baseline-derivation-attack}) extra receipts parsing.
\end{enumerate}

\begin{lstlisting}[language=Python,caption={Prefeed and bloom gates (pseudocode)}]
# Gate 1: DA-aware prefeeding
function next_data(l1_block):
    if l1_block.number not in B: return EOF  # skip non-DA blocks entirely
    return extract_batch_data(l1_block)  # authenticate sender, 'to', blobs

# Gate 2: bloom-gated SystemConfig updates
function maybe_update_system_config(l1_block):
    if not bloom_may_contain(l1_block.header, CONFIG_UPDATE_TOPIC): return
    receipts = fetch_receipts(l1_block.hash)
    system_config.apply(receipts)  # may change batcher EOA and other params
\end{lstlisting}

Prefeeding means the zkVM does not scan skipped blocks; therefore we need a \emph{cryptographic invariant} to ensure no DA-bearing batcher transaction was omitted or reordered. The batcher EOA’s nonce provides exactly this:
\[
\text{for consecutive batcher txs } (t_i),\quad \mathrm{nonce}(t_{i+1})=\mathrm{nonce}(t_i)+1.
\]
A gap or regression \emph{must} surface as a nonce discontinuity. The zkVM enforces this invariant over all batcher txs that it ingests in the range.
As for the starting point of nonce tracking, we assert that the initial tracked nonce is less than or equal to the agreed-upon nonce.

\begin{lstlisting}[language=Python,caption={Nonce discipline inside extraction (pseudocode)}]
first_tx = batcher_txs_in_range[0]
assert first_tx.sender == agreed_sender
assert first_tx.nonce <= agreed_nonce 

expected_sender, expected_nonce = agreed_sender, first_tx.nonce

for tx in batcher_txs_in_range[1:]:
    assert tx.nonce == expected_nonce
    assert tx.sender == expected_sender
    expected_nonce = tx.nonce + 1 # strict + 1, no gaps, no dupes
    expected_sender = tx.sender
# (expected_sender, expected_nonce) becomes the range's post state
\end{lstlisting}

\subsection*{Why batcher changes complicate the approach (and when we fall back to baseline)}
\icode{SystemConfig} can update the batcher EOA. At the exact L1 block of a change: (i) the expected sender switches, (ii) the nonce sequence restarts under the new sender, and (iii) DA must be authenticated against the \emph{updated} \icode{SystemConfig} for that block. Missing this update would make prefeed+nonce unsound, so we (a) \emph{detect} changes before derivation, (b) \emph{wire} the pipeline accordingly, (c) \emph{authenticate} per block under the current config, and (d) \emph{verify} postconditions.

\paragraph{Detection (pre-pass, bloom-gated).}
Only on bloom hits do we fetch/parse receipts; any batcher EOA change disables optimizations resorting to the ``baseline'' derivation.
\begin{lstlisting}[language=Python,caption={Bloom-gated detection of batcher EOA change}]
function has_batcher_sender_change(origin, head) -> bool:
    sys = system_config_at_L2_safe_head(origin)
    b   = origin
    while b.number < head.number:
        hdr = l1.header_by_hash(b.hash)
        if bloom_may_contain(hdr, CONFIG_UPDATE_TOPIC): 
            rxs = l1.receipts_by_hash(b.hash)
            old = sys.batcher_address
            sys.apply_receipts(rxs)                 # may update batcher EOA
            if sys.batcher_address != old: return true
        b = l1.block_info_by_number(b.number + 1)
    return false
\end{lstlisting}

\paragraph{Wiring (optimized vs.\ baseline).}
If a change is detected anywhere in \([{\tt origin},{\tt head}]\), use \emph{baseline} for that run; otherwise enable prefeed+nonce.
\begin{lstlisting}[language=Python,caption={Pipeline configuration based on detection outcome}]
change = has_batcher_sender_change(origin, head)

if change == false:
    agreed = (boot.agreed_sender_address, boot.agreed_nonce)
    da_provider = new_EthereumDataSource(l1, beacon, cfg, agreed)   # enable nonce tracking
    B = l1.prefed_DA_block_numbers(origin)                          # sparse DA blocks
    pipeline   = new_OraclePipeline(cfg, cursor, oracle, da_provider, l1, l2, B)
else:
    da_provider = new_EthereumDataSource(l1, beacon, cfg, None)     # no hints
    pipeline   = new_OraclePipeline(cfg, cursor, oracle, da_provider, l1, l2, None)
\end{lstlisting}

\paragraph{Per-block checks}
Update \icode{SystemConfig} for block \(h\) \emph{before} accepting any DA from \(h\); authenticate batcher txs against the \emph{current} sender; process DA content.
\begin{lstlisting}[language=Python,caption={Per-block: update SystemConfig, then process DA}]
for h in L1_interval: 
    if is_DA_block(h):                           # true only in optimized mode with B
        if bloom_may_contain(header(h), CONFIG_UPDATE_TOPIC):
            rxs = l1.receipts_by_hash(h.hash)
            sys.apply_receipts(rxs)                  # pick up batcher change at h
    
            for tx in batcher_txs(h):                # L1 order
                assert tx.from == sys.batcher_address_at(h)
                assert (calldata -> tx.to == BATCH_INBOX) or (blob -> check_blob_hashes(tx))
                process_da_bytes(tx)
\end{lstlisting}

\paragraph{Epilogue checks}
Enforce the claimed boundary \((\texttt{sender},\texttt{nonce})\).
\begin{lstlisting}[language=Python,caption={Post-range boundary verification}]
(post_sender, post_nonce) = da_provider.tracked_nonce_state()
assert post_sender == boot.claimed_sender_address
assert post_nonce  == boot.claimed_nonce
\end{lstlisting}

\subsection*{How noncing applies at \emph{range} and \emph{aggregate} levels}
\paragraph{Range.} A range proof validates derivation over a contiguous L1 interval. It enforces:
\[
\resizebox{\textwidth}{!}{$
(\texttt{sender}, \texttt{nonce}) \text{ evolves by }
\begin{cases}
\texttt{sender}'=\texttt{sender},\ \ \texttt{nonce}'=\texttt{nonce}+1 & \text{(next batcher tx)}\\
\texttt{sender}'=\texttt{sender},\ \ \texttt{nonce}'=\texttt{nonce}   & \text{(no DA in this block)}\\
\texttt{sender}'=\texttt{newBatcher},\ \ \texttt{nonce}'=\texttt{reset} & \text{(batcher change)}
\end{cases}
$}
\]
and rejects any deviation. If a batcher change occurs \emph{inside} the interval, the proof explicitly shows the boundary and applies the updated \icode{SystemConfig} before further DA acceptance.

Each accepted batcher transaction is checked against the \emph{expected sender} and a strictly incrementing \emph{expected nonce}. Two error cases are explicitly enforced:
\begin{itemize}
  \item \textbf{Nonce gap error:} if \icode{tx.nonce > state.nonce + 1}, a transaction was skipped.
  \item \textbf{Nonce mismatch error:} if \icode{tx.nonce != state.nonce + 1}, then ordering or replay has been violated.
\end{itemize}
These checks guarantee that no DA transaction can be silently omitted or reordered: any deviation in the nonce chain will immediately cause the proof to reject.

\begin{lstlisting}[language=Python,caption={Range program: nonce/sender evolution with errors (pseudocode)}]
state = (start_sender, start_nonce)

for h in L1_interval:
    if h in B:                               # DA-bearing L1 block
        for tx in batcher_txs(h):            # L1 order
            if tx.sender != state.sender:
                raise SenderMismatchError
            if tx.nonce != state.nonce:
                if tx.nonce > state.nonce:
                    raise NonceGapError      # skipped transaction(s)
                else:
                    raise NonceMismatchError # replay or reordering
            state.nonce = tx.nonce + 1
# output post_state = state
\end{lstlisting}

\paragraph{Aggregate.} Aggregation composes multiple ranges. It requires \emph{boundary alignment}:
\[
\texttt{sender}_\text{end}^{(i)}=\texttt{sender}_\text{start}^{(i+1)},\quad
\texttt{nonce}_\text{end}^{(i)}=\texttt{nonce}_\text{start}^{(i+1)}.
\]
This ensures the nonce/sender invariant extends across the entire aggregated trace.

\begin{lstlisting}[language=Python,caption={Aggregate program: cross-range continuity (pseudocode)}]
for i in 0..k-1:
    assert ranges[i].post_sender == ranges[i+1].pre_sender
    assert ranges[i].post_nonce  == ranges[i+1].pre_nonce
\end{lstlisting}

\subsection*{Chaining and L1 anchoring (contract changes)}
To bind range and aggregate proofs to L1 ground truth, we extend \icode{L2OutputOracle} with batched metadata and enforce it at proposal time.

\begin{lstlisting}[language=Python,caption={L1 contract anchoring (pseudocode)}]
struct OutputMeta { address batcher; uint64 nonce; }
OutputMeta meta;   // last accepted (batcher, nonce)

function proposeOutputRoot(root, proof, pre_sender, pre_nonce, post_sender, post_nonce):
    require(pre_sender == meta.batcher && pre_nonce == meta.nonce)
    verify_aggregate_proof(proof, /* includes range proofs and DA commitments */)
    meta.batcher = post_sender
    meta.nonce   = post_nonce
    store_output_root(root)
\end{lstlisting}

\noindent
\emph{Anchoring.} Each range proof commits to its \icode{(pre_sender, pre_nonce)} and \icode{(post_sender, post_nonce)}; the aggregate proof checks cross-range alignment, and \icode{L2OutputOracle} checks alignment with on-chain \icode{meta}. Thus, host-side prefeeding cannot suppress DA: a missing batcher tx would break the nonce chain either within a range or at an aggregation boundary, and the contract would reject the submission.

\subsection*{Edge cases (and mixed aggregation)}
\begin{itemize}
 \item \textbf{Batcher change within a range.}  
If a batcher EOA change is detected anywhere between the range’s origin and head, the optimized pipeline is \emph{disabled} and the range is derived in baseline mode. This means no DA-prefeeding and nonce continuity is no longer enforced; instead, the full L1 traversal with receipt parsing is executed. The range proof then reports its \icode{pre_sender, pre_nonce} from the start L1 head and its \icode{post_sender, post_nonce} from the end L1 head, exactly as returned by:
\begin{verbatim}
let (agreed_sender, agreed_nonce) =
    get_batcher_sender_info_at(start_l1_head);
let (claimed_sender, claimed_nonce) =
    get_batcher_sender_info_at(end_l1_head);
\end{verbatim}
This ensures that any batcher reset is faithfully reflected in the range boundaries, and aggregation simply enforces equality of these boundary pairs.

  \item \textbf{Optimized + baseline mixing.}  
  Aggregation does not care how an individual range was derived. Ranges produced by the optimized pipeline (DA-prefeed + nonce enforcement) and those from the baseline pipeline (full L1 scan) can be freely aggregated. The only condition is boundary equality: the \icode{post_sender, post_nonce} of one range must match the \icode{pre_sender, pre_nonce} of the next. The on-chain contract checks only these boundary values, so the internal derivation method is irrelevant to soundness.

  \item \textbf{Bloom false positives.}  
  Bloom filters may over-approximate configuration changes, causing extra receipt fetches. This does not affect safety. Because blooms have no false negatives for topics, genuine batcher EOA updates cannot be missed (see ~\Cref{sec:baseline-derivation-attack} for an in depth analysis).

\item \textbf{Aggregation at batcher-change boundaries.}  
Each range is anchored by two L1 heads:
\begin{itemize}
  \item the \emph{start L1 head}, returned by \icode{get_l1_head(l2_start_block)}, and
  \item the \emph{end L1 head}, returned by \icode{get_l1_head(l2_end_block)}.
\end{itemize}
These are not arbitrary counters: they are the actual L1 blocks from which the L2 start and end blocks can be derived.
At each anchor, the batcher state is read directly from L1:
\begin{verbatim}
let (sender, nonce) = get_batcher_sender_info_at(l1_head);
\end{verbatim}

\textbf{Aggregation rule.}  
For consecutive ranges, aggregation enforces
\[
\text{Range}_i.\,\texttt{postNonce} \;=\; \text{Range}_{i+1}.\,\texttt{preNonce}.
\]
Since $\text{Range}_i.\text{end L1 head} = \text{Range}_{i+1}.\text{start L1 head}$, both ranges read their boundary state from the \emph{same} L1 block.  
This guarantees consistent values, even across batcher changes.

\textbf{Example.}  
A batcher change occurs at L1\#120: the old batcher had reached nonce 200, and the new batcher begins with nonce 1.
\begin{itemize}
  \item Range$_1$: start L1 head = L1\#100, end L1 head = L1\#120.  
        At L1\#120, \icode{post_sender, post_nonce} = (new batcher, 1).
  \item Range$_2$: start L1 head = L1\#120, end L1 head = L1\#130.  
        At the same L1\#120, \icode{pre_sender, pre_nonce} = (new batcher, 1).
\end{itemize}

Thus both sides of the boundary agree on the identical pair (new batcher, nonce 1).  
There is no possibility of a mismatch like “200 vs 1,” because the overlap forces both to use the same on-chain state.

This construction ensures that range aggregation remains valid across batcher-change epochs, since the boundary state is anchored in L1 and shared.
\end{itemize}

The application inside zkVM never searches L1 for DA nor parses receipts indiscriminately. Instead, it processes (i) a sparse sequence of DA-bearing blocks and (ii) the rarer bloom-hit blocks for configuration changes. Trustlessness is preserved by enforcing a global nonce/sender invariant at both \emph{range} and \emph{aggregate} levels, and by anchoring boundary state on L1 via \icode{L2OutputOracle}. This yields substantial proving savings without altering Optimism’s safety or execution semantics.

\section{Evaluation}\label{sec:evaluation}
\subsection{Benchmark Methodology and Setup}
To quantify the impact of our optimized derivation pipeline, we benchmarked both the \emph{baseline} (unoptimized Kona fault-proof program) and the \emph{optimized} variant incorporating nonce-based prefeeding and bloom-gated derivation.  
All experiments use SP1’s built-in \emph{cycle counter} as the primary cost metric. 
SP1 provides a cycle-tracking API that reports the total number of RISC-V instruction cycles 
executed by a program~\cite{sp1-docs}. 
Empirically, we observe that total cycle counts scale \emph{linearly} with wall-clock proof generation time, 
making cycle count a faithful proxy for prover effort. 
We therefore adopt cycle count as the canonical measure of computational cost throughout this evaluation.

\paragraph{Metrics.}
To understand where our optimizations yield benefits, we decompose total zkVM execution into three measurable components, each directly tied to one of our optimization objectives:
\begin{itemize}
  \item \textbf{Total Instruction Count:} The aggregate number of SP1 execution cycles, capturing the full prover workload across derivation, execution, and verification. This serves as our primary measure of overall computational cost.
  \item \textbf{Derivation Count:} The subset of cycles consumed by the L1 traversal and data-availability (DA) extraction pipeline. 
        Reductions here quantify the impact of our \emph{sparse prefeeding} optimization, which avoids scanning non–DA-bearing L1 blocks.
  \item \textbf{Receipt Count:} The cycles attributed to decoding and applying \icode|SystemConfig| updates from L1 receipts.
        These are expected to decrease proportionally with our \emph{bloom-gated} receipt filtering.
\end{itemize}

\paragraph{Experimental design.}
We evaluate our optimizations through three controlled experiments designed to isolate the main cost drivers of the derivation pipeline. Across all experiments, we fix the \emph{range size} at \textbf{400} L2 blocks, allowing each proof to cover the same span of L2 state regardless of the underlying L1 derivation workload.

\begin{itemize}
  \item \textbf{Experiment 1 — Scaling with channel timeout.}  
  The derivation workload within the zkVM scales with the number of L1 blocks that must be processed per proof.  
  To measure this relationship, we vary the \emph{channel timeout} parameter—the maximum permissible interval between consecutive batch submissions.  
  Larger timeouts cause more unbatched L1 blocks to accumulate, thereby increasing baseline cost and highlighting how performance scales with derivation depth.

  \item \textbf{Experiment 2 — Bloom false positives.}  
  This experiment examines the sensitivity of the optimized pipeline to worst-case bloom filter behavior.  
  A synthetic $100\%$ false-positive rate forces every block to fetch receipts, effectively disabling bloom gating.  
  This setup quantifies the upper bound on wasted work and demonstrates that, even in pathological conditions, correctness and substantial efficiency gains are preserved.

  \item \textbf{Experiment 3 — Batcher outage.}  
  Finally, we assess resilience under real-world operational stress.  
  The \texttt{op-batcher} is critical for feeding DA-bearing transactions; when it stalls, the derivation pipeline must later traverse a large L1 backlog.  
  To capture this scenario, we replay a \emph{historical 2.5-hour outage} observed in mainnet data and measure total zkVM cycles during and after recovery.  
  This test evaluates how well the optimized derivation absorbs delayed data availability and how quickly it returns to steady-state performance.
\end{itemize}

Across all experiments, we record three primary metrics—total instruction count, derivation count, and receipt count (as defined earlier)—to measure both aggregate and component-level improvements in zkVM computation.

\subsection{Results}

\subsubsection{Scaling with channel timeout (Baseline vs.\ Optimized.)}
Table~\ref{tab:baseline-optimized} summarizes the results across multiple \emph{channel timeout} configurations.  
As the timeout increases, the number of L1 blocks included in each proof grows, leading to higher total workload for the baseline pipeline.  
The optimized pipeline exhibits significant reductions across all metrics, with the largest gains observed in \emph{Derivation Count} (up to $84\%$ improvement) and \emph{Receipt Count} (over $84\%$ improvement).  
These directly correspond to our two main optimizations: sparse prefeeding of DA-bearing blocks and bloom-gated \icode{SystemConfig} updates.

\begin{table}[h]
\centering
\caption{Comparison between Baselin (Bse.) and Optimized (Opt.) pipelines. Improvements (Impr.\%) are relative reductions in cycle counts.}
\label{tab:baseline-optimized}
\makebox[0pt][c]{%
  \resizebox{1\linewidth}{!}{%
\begin{tabular}{ccc ccc ccc ccc}
\toprule
\multirow[b]{2}{*}{Channel (s)} & \multirow[b]{2}{*}{L1 Blocks} & \multirow[b]{2}{*}{L1 Txs}
  & \multicolumn{3}{c}{Total Instruction Count}
  & \multicolumn{3}{c}{Derivation Count}
  & \multicolumn{3}{c}{Receipt Count} \\
\cmidrule(lr){4-6}\cmidrule(lr){7-9}\cmidrule(lr){10-12}
 &  & 
 & Bse. & Opt. & Impr. \%
 & Bse. & Opt. & Impr. \%
 & Bse. & Opt. & Impr. \% \\
\midrule
10  & 123 & 23,576 & 3.26B & 2.49B & 23.49 & 1.84B & 1.16B & 37.17 & 0.68B & 0.29B & 56.57 \\
50  & 145 & 27,727 & 2.85B & 1.62B & 43.09 & 1.63B & 0.53B & 67.65 & 0.75B & 0.25B & 66.85 \\
100 & 174 & 33,205 & 3.05B & 1.52B & 50.12 & 1.83B & 0.47B & 74.13 & 0.86B & 0.25B & 71.24 \\
200 & 235 & 44,883 & 3.64B & 1.51B & 58.55 & 2.31B & 0.45B & 80.67 & 1.08B & 0.24B & 77.44 \\
500 & 382 & 72,950 & 4.93B & 1.40B & 71.63 & 3.59B & 0.55B & 84.75 & 1.61B & 0.25B & 84.52 \\
\bottomrule
\end{tabular}
  }
}
\end{table}

\subsubsection{Bloom False Positives.}\label{sec:baseline-derivation-attack}
Table~\ref{tab:falsepositives} isolates the pathological scenario where the bloom filter produces a $100\%$ false-positive rate, forcing every block to fetch receipts.  
Even in this worst case, the optimized pipeline maintains measurable savings, since sparse prefeeding continues to eliminate redundant DA traversal.  
In realistic settings, bloom false-positive rates are typically below $1\%$, meaning that the actual performance gap is expected to align closely with Table~\ref{tab:baseline-optimized}.

\begin{table}[h]
\centering
\caption{Comparison between Baseline (Bse.) and Optimized (Opt.) with 100\% false-positive rate. Improvements (Impr.\%) are relative reductions in cycle counts.}\label{tab:falsepositives}
\makebox[0pt][c]{%
  \resizebox{1\linewidth}{!}{%
\begin{tabular}{ccc ccc ccc ccc}
\toprule
\multirow[b]{2}{*}{Channel (s)} & \multirow[b]{2}{*}{L1 Blocks} & \multirow[b]{2}{*}{L1 Txs}
  & \multicolumn{3}{c}{Total Instruction Count}
  & \multicolumn{3}{c}{Derivation Count}
  & \multicolumn{3}{c}{Receipt Count} \\
\cmidrule(lr){4-6}\cmidrule(lr){7-9}\cmidrule(lr){10-12}
 &  & 
 & Bse. & Opt. & Impr. \%
 & Bse. & Opt. & Impr.\%
 & Bse. & Opt. & Impr.\% \\
\midrule
10  & 123 & 23,576 & 3.26B & 2.87B & 11.76 & 1.84B & 1.54B & 16.45 & 0.68B & 0.68B & 0.00 \\
50  & 145 & 27,727 & 2.85B & 2.13B & 25.43 & 1.63B & 1.03B & 36.73 & 0.75B & 0.75B & 0.00 \\
100 & 174 & 33,205 & 3.05B & 2.13B & 30.12 & 1.83B & 1.08B & 40.81 & 0.86B & 0.86B & 0.00 \\
200 & 235 & 44,883 & 3.64B & 2.34B & 35.63 & 2.31B & 1.28B & 44.57 & 1.08B & 1.08B & 0.00 \\
500 & 382 & 72,950 & 4.93B & 2.76B & 44.12 & 3.59B & 1.91B & 46.94 & 1.61B & 1.61B & 0.00 \\
\bottomrule
\end{tabular}
  }
}
\end{table}

Because the false-positive rate of the logs bloom filter scales with the number of logs per block, we should expect more false positives as the log volume of the block grows. The draft EIP-7745~\cite{eip7745} proposes replacing the header bloom with an adaptive data structure whose capacity scales with per-block log counts, thereby maintaining a consistently low false positive rate. Note that EIP-7745 is complementary to our work: while we optimize derivation by using the existing bloom filter to gate receipt fetching, EIP-7745 would improve the bloom filter itself to reduce false positives. Adopting EIP-7745 would enhance our optimization's effectiveness but does not overlap with our core contributions of nonce-based sparse prefeeding and range aggregation.

\subsubsection{Batcher outage.}
\begin{figure}[htb!]
  \centering
  \resizebox{\linewidth}{!}{\input{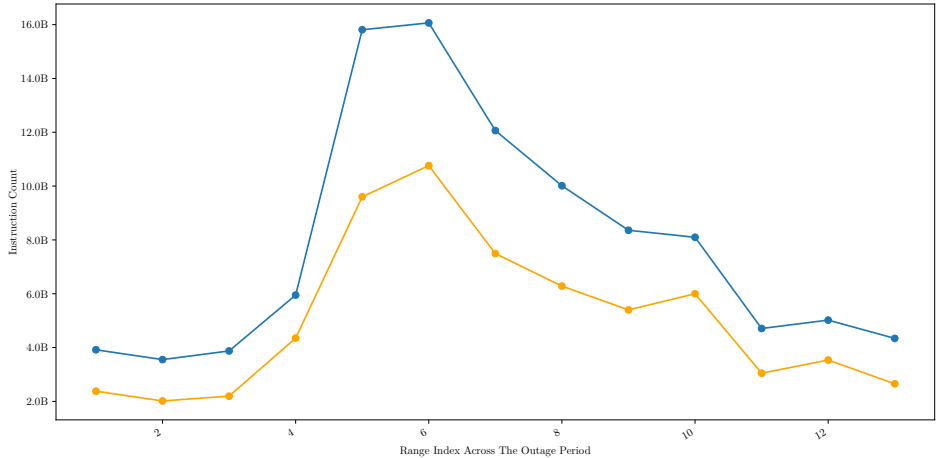}}
  \caption{Impact of a temporary batcher outage (for two and half hours) on total instruction count. 
  Both pipelines spike during backlog catch-up; the optimized pipeline tracks the same shape but with a lower peak (\(\sim\)10B vs.\ \(\sim\)16B cycles at the maximum) and a faster post-outage decay.}
  \label{fig:outage}
\end{figure}

We simulate a temporary \texttt{op-batcher} outage lasting two and a half hours —roughly 
\(\tfrac{2.5 \times 3600}{12} \approx 750\) Ethereum L1 blocks. 
When the batcher resumes posting, the derivation pipeline must replay this entire backlog to bring the L2 tip back in sync with the L1 head. 
Figure~\ref{fig:outage} shows that both the baseline and optimized pipelines exhibit a pronounced spike in cycle count during this catch-up phase, proportional to the accumulated L1 work, but the optimized pipeline consistently incurs 50–60\% fewer cycles at peak.

After a batcher outage, there’s a large DA backlog but the batcher can only resume posting at L1’s bounded throughput (gas and 4844‑blob limits per block), so that backlog is spread across many subsequent L1 blocks/channels. Derivation is strictly sequential and range‑bounded: the first post‑outage range can only consume the frames that were actually posted between its start/end L1 heads, while the remaining backlog lands in later blocks and must be processed by later ranges. Consequently, cycles stay elevated for several ranges—until the backlog is fully posted and ingested—rather than snapping back to normal after one.

\subsection{Security Analysis}

Having established the performance benefits of our optimized derivation pipeline, we now turn to its security properties. Our analysis addresses two complementary concerns: (1) whether the optimizations preserve the soundness guarantees of the baseline protocol, and (2) whether adversaries can exploit the optimization mechanisms to degrade performance. We demonstrate that the design maintains provable soundness through explicit defense mechanisms while remaining robust against targeted performance attacks.

\subsubsection{Soundness Evaluation}

The primary security requirement for any fault-proof system is \emph{derivation soundness}: the prover must produce outputs that faithfully reflect the canonical L2 state derived from L1 data availability, and any deviation must be detectable and rejectable. Our optimizations—nonce-based sparse prefeeding and Bloom-gated receipt parsing—introduce new code paths that could potentially admit forgery, omission, or reordering attacks if not carefully designed.

To establish soundness, we first enumerate the security-critical invariants enforced by our design, then systematically analyze potential attack vectors using an attack–defense tree formalism.

\paragraph{Soundness invariants.}
The optimized derivation pipeline enforces three critical invariants that collectively ensure soundness:

\begin{enumerate}
  \item \textbf{Nonce continuity.} Every data-availability transaction from the active batcher must increment the sender nonce by exactly one. Nonce gaps or regressions trigger immediate rejection via \texttt{NonceGapError}.
  
  \item \textbf{Batcher authenticity.} Only transactions cryptographically signed by the address registered in \texttt{SystemConfig.}\allowbreak\texttt{batcherAddress} are accepted as valid DA sources. Signature verification occurs per-transaction and raises \texttt{SenderMismatchError} on mismatch.
  
  \item \textbf{Cross-range consistency.} Adjacent proof ranges must maintain (sender, nonce) continuity at their boundaries. The \texttt{L2OutputOracle} anchors these boundary states in L1, enabling the aggregator to detect any inconsistency via \texttt{NonceMismatchError}.
\end{enumerate}

These invariants are sufficient to prevent an adversary from injecting forged data, omitting legitimate batches, or reordering the canonical sequence without detection.

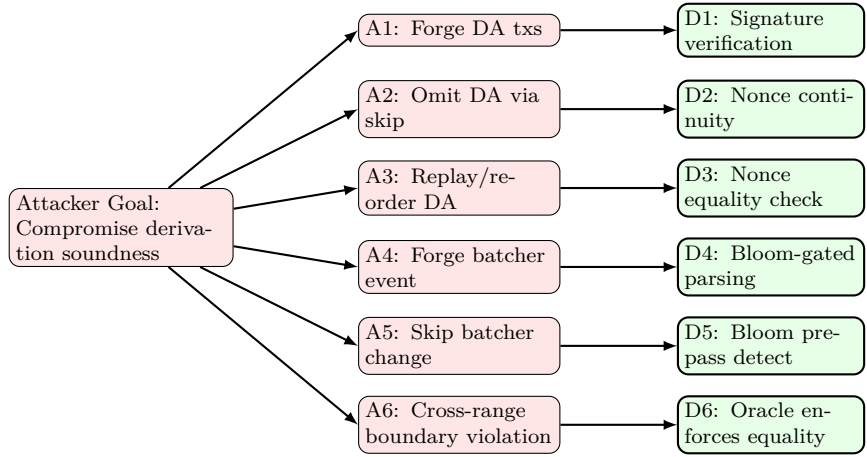
\begin{figure}[htb!]
\centering
\scriptsize
\begin{tikzpicture}[
  grow=right,
  growth parent anchor=east,
  child anchor=west,
  level 1/.style={level distance=3cm, sibling distance=1.05cm},
  level 2/.style={level distance=2.8cm, sibling distance=1.05cm},
  every node/.style={draw, rounded corners, align=left, font=\scriptsize, inner sep=2.5pt},
  edge from parent/.style={draw, -latex, thick},
  atk/.style={fill=red!10, text width=2.5cm},
  def/.style={fill=green!10, text width=2.3cm}
]
\node[atk, text width=2.8cm]{Attacker Goal:\\Compromise derivation soundness}
  child { node[atk]{A6: Cross-range boundary violation}
    child { node[def]{D6: Oracle enforces equality} }
  }
  child { node[atk]{A5: Skip batcher change}
    child { node[def]{D5: Bloom pre-pass detect} }
  }
  child { node[atk]{A4: Forge batcher event}
    child { node[def]{D4: Bloom-gated parsing} }
  }
  child { node[atk]{A3: Replay/reorder DA}
    child { node[def]{D3: Nonce equality check} }
  }
  child { node[atk]{A2: Omit DA via skip}
    child { node[def]{D2: Nonce continuity} }
  }
  child { node[atk]{A1: Forge DA txs}
    child { node[def]{D1: Signature verification} }
  };
\end{tikzpicture}
\caption{Attack–defense tree for the optimized derivation pipeline. Each attack vector (red) is countered by a defense mechanism (green). Defenses: D1 raises \texttt{SenderMismatchError}; D2 raises \texttt{NonceGapError}; D3 raises \texttt{NonceMismatchError}; D4 authenticates \texttt{SystemConfig} logs; D5 ensures completeness via baseline fallback; D6 validates (sender, nonce) at aggregation boundaries.}
\label{fig:attack-tree-final}
\end{figure}

\paragraph{Attack–defense analysis.}
Figure~\ref{fig:attack-tree-final} presents a systematic enumeration of attack vectors and their corresponding defenses. We organize threats into six categories:

\begin{itemize}
  \item \textbf{A1: Forge DA transactions.} An attacker attempts to inject transactions that appear to originate from the legitimate batcher. \emph{Defense:} Per-transaction ECDSA signature verification ensures that only transactions signed by \texttt{SystemConfig.}\allowbreak\texttt{batcherAddress} are accepted (D1). Any forgery triggers \texttt{SenderMismatchError}.
  
  \item \textbf{A2: Omit DA via prefeed skip.} The sparse prefeeding optimization could be exploited to skip blocks containing legitimate DA. \emph{Defense:} Nonce continuity enforcement guarantees that any omitted transaction creates a detectable gap in the nonce sequence (D2), raising \texttt{NonceGapError}.
  
  \item \textbf{A3: Replay or reorder DA.} An attacker might attempt to reuse or permute previously posted batches. \emph{Defense:} Strict nonce equality checks prevent both replay (same nonce reused) and reordering (nonce sequence violated), triggering \texttt{NonceMismatchError} (D3).
  
  \item \textbf{A4: Forge batcher-change event.} An adversary could inject a fake \texttt{SystemConfig} update to redirect DA acceptance to a compromised address. \emph{Defense:} Bloom-gated receipt parsing first filters candidates using the Ethereum header Bloom filter, then cryptographically validates that any \texttt{ConfigUpdate} event originates from the authentic \texttt{SystemConfig} contract (D4).
  
  \item \textbf{A5: Skip legitimate batcher change.} The Bloom gating could be circumvented to omit genuine batcher updates. \emph{Defense:} The two-pass Bloom architecture ensures completeness: if the optimized path produces fewer results than the baseline scan, the system falls back to exhaustive receipt parsing (D5).
  
  \item \textbf{A6: Cross-range boundary violation.} An attacker might attempt to forge continuity at the boundary between adjacent proof ranges. \emph{Defense:} The \texttt{L2OutputOracle} anchors boundary (sender, nonce) pairs in L1 state, enabling the aggregator to enforce strict equality across range transitions (D6).
\end{itemize}

Each defense mechanism is implemented as an explicit check in the zkVM execution trace, ensuring that violations produce deterministic errors visible to verifiers. The combination of cryptographic authentication (signature verification), structural constraints (nonce continuity), and L1 anchoring (oracle-enforced boundaries) provides defense-in-depth against the identified attack surface.

\subsubsection{Performance Degradation Resistance}

While the soundness analysis establishes that optimizations cannot be exploited to violate correctness, a separate concern is whether adversaries can \emph{degrade performance} by forcing the optimized pipeline to behave like the baseline. In ~\Cref{tab:falsepositives}, we empirically demonstrated that even a pathological 100\% Bloom false-positive rate maintains substantial efficiency gains due to sparse prefeeding. Here we analyze whether such conditions can be deliberately induced.

\paragraph{Threat model.}
We consider an adversary who can submit arbitrary transactions to Ethereum L1 with the goal of maximizing false positives in the Bloom filter used for \texttt{SystemConfig} event detection. Success would force the derivation pipeline to fetch and decode receipts for every L1 block, effectively disabling Bloom gating.

\paragraph{Bloom filter mechanics.}
Ethereum's header Bloom filter is a 2048-bit ($m = 2048$) probabilistic data structure. For each log, the emitting contract address and each indexed topic are hashed via Keccak-256, producing three bit positions $p_1, p_2, p_3 \in \{0, \ldots, 2047\}$ ($k = 3$) that are set in the filter. A query matches if all its required bits are set.

\paragraph{Attack construction.}
Because the Bloom filter aggregates bits via bitwise OR, an adversary can construct collisions by emitting logs with carefully chosen topics:

\begin{enumerate}
  \item \textbf{Topic collision.} If the victim query specifies $m$ topics (e.g., event signature plus indexed parameters), the attacker emits logs whose topics are \emph{exactly} those $m$ values. This deterministically sets the $3m$ topic-derived bits.
  
  \item \textbf{Address bit coverage.} The three bits derived from \texttt{SystemConfig}'s address must also be set. Rather than searching for a single topic whose three bits exactly match the address (probability $\approx 6.98 \times 10^{-10}$), the attacker can \emph{cover} these bits across multiple topics. Since a random topic includes any specific target bit with probability $3/2048$, covering all three address bits requires approximately $3 \times (2048/3) \approx 2048$ topic candidates—a trivial brute-force search.
\end{enumerate}

\paragraph{Gas cost analysis.}
Emitting a log with $T$ topics and $D$ data bytes costs:
\[
  \text{Gas}_{\text{LOG}}(T, D) \approx 375 + 375T + 8D.
\]
For a query with $m$ topics, the attacker needs $m + 3$ total topics (query topics plus address coverage). Since EVM logs support at most 4 topics, this requires $\lceil (m+3)/4 \rceil$ logs. With $D = 0$, the gas cost is:
\[
  \text{Gas}_{\text{attack}} \approx 375 \cdot \left( (m+3) + \left\lceil \frac{m+3}{4} \right\rceil \right),
\]
which is modest relative to Ethereum's per-block gas limit (typically $\sim$30M gas).

\paragraph{Impact assessment.}
While this attack can force Bloom false positives, its practical impact is limited by three factors:

\begin{enumerate}
  \item \textbf{Residual optimization.} As demonstrated in Table~\ref{tab:falsepositives}, sparse prefeeding continues to provide 25–43\% cycle reduction even under 100\% false-positive conditions. The attack degrades Bloom gating but cannot eliminate the nonce-based prefeed optimization.
  
  \item \textbf{Economic cost.} Sustained attacks require continuous L1 gas expenditure. At current gas prices, manufacturing false positives across every block would impose significant recurring costs on the attacker while providing only marginal degradation.
  
  \item \textbf{Mitigation path.} The draft EIP-7745~\cite{eip7745} proposes replacing the fixed-size Bloom filter with an adaptive data structure whose capacity scales with per-block log volume, maintaining consistently low false-positive rates even under adversarial conditions.
\end{enumerate}

In summary, while targeted Bloom attacks are theoretically feasible, they impose greater costs on attackers than defenders and fail to fully negate the optimization benefits. The design remains robust against realistic performance degradation threats.
\section{Conclusions}
We presented a soundness-preserving optimization of Optimism's fault-proof derivation that reduces zkVM proving costs by up to $3.5\times$ overall and $6.5\times$ in derivation. Through nonce-authenticated sparse prefeeding, Bloom-gated receipt parsing, and L1-anchored range aggregation, our approach eliminates redundant L1 traversal and receipt processing—the dominant cost drivers in zkVM-based rollup derivation. Experimental evaluation demonstrates ~85\% reduction in derivation cycles and ~72\% lower total instruction count under high channel timeouts, with gains persisting even under adversarial conditions (100\% Bloom false-positive rate) and operational stress (2.5-hour batcher outage). Security analysis confirms that nonce continuity enforcement, signature verification, and on-chain boundary anchoring prevent forgery, omission, and reordering attacks without weakening protocol guarantees.

More broadly, our work demonstrates that \emph{verifiable execution does not imply efficient execution}. While zkVMs enable trustless verification of arbitrary computation, co-designing protocols with proving constraints in mind—minimizing input traversal, amortizing cryptographic checks, and anchoring invariants on-chain—yields disproportionate performance gains. This principle extends beyond rollup derivation to any zkVM application processing large, partially-relevant input streams. Future work includes extending these optimizations to deposit processing, evaluating portability across alternative zkVMs, and developing formal verification frameworks for derivation soundness.
\bibliographystyle{plain}
\bibliography{references}

\end{document}